\documentstyle[epsfig,aps,prb]{revtex}
\newcommand\beq{\begin{equation}}
\newcommand\eeq{\end{equation}}
\newcommand\bea{\begin{eqnarray}}
\newcommand\eea{\end{eqnarray}}

\begin{document}

\begin{center}
{\Large Evaluation of low-energy effective Hamiltonian techniques for
coupled spin triangles}
\end{center}

\vskip .5 true cm
\centerline{\bf C. Raghu$^1$, Indranil Rudra$^1$, S. Ramasesha$^1$}
\centerline{\bf and Diptiman Sen$^2$} 
\vskip .5 true cm

\centerline{\it $^1$ Solid State and Structural Chemistry Unit} 
\centerline{\it $^2$ Centre for Theoretical Studies}  
\centerline{\it Indian Institute of Science, Bangalore 560012, India} 
\vskip .5 true cm

\begin{abstract}

Motivated by recent work on Heisenberg antiferromagnetic
spin systems on various lattices made up of triangles, we examine the 
low-energy properties of a chain of antiferromagnetically coupled triangles 
of half-odd-integer spins. We derive the low-energy effective Hamiltonian to 
second order in the ratio of the coupling $J_2$ between triangles to the 
coupling $J_1$ within each triangle. The effective Hamiltonian contains 
four states for each triangle which are given by the products of spin-$1/2$ 
states with the states of a pseudospin-$1/2$. We compare the
results obtained by exact diagonalization of the effective Hamiltonian with 
those obtained for the full Hamiltonian using exact diagonalization
and the density-matrix renormalization group method. It is found that the
effective Hamiltonian is accurate only for the ground state for rather low 
values of the ratio $J_2 / J_1$ and that too for the spin-$1/2$ case with 
linear topology. The chain of spin-$1/2$ triangles shows interesting 
properties like spontaneous dimerization and several singlet and triplet 
excited states lying close to the ground state.

\end{abstract}
\vskip .5 true cm

~~~~~~ PACS number: ~75.10.Jm, ~75.50.Ee

\newpage

\section{Introduction}

Low-dimensional quantum spin systems with frustration (i.e., competing 
antiferromagnetic interactions) have been studied extensively in recent 
years. A particularly interesting class of such systems have lattice 
structures built out of triangles of spins. Many such systems
have been realized experimentally in one and two dimensions, such
as the sawtooth chain \cite{saw} and the Kagome lattice \cite{kag1}.
A variety of theoretical techniques, both analytical and numerical, are 
available to study the relevant models \cite{kag2,kag3,lech,kagstrip}. Due to 
large quantum fluctuations, such systems often have several unusual 
properties. For instance, there may be 
a gap between a singlet ground state and the nonsinglet excited states; this 
leads to a magnetic susceptibility which vanishes exponentially at 
low temperatures. Secondly, the two-spin correlation function may
decay rapidly with distance indicating that the magnetic correlation
length is not much bigger than the lattice spacing. Finally, there is 
sometimes no gap to a large number of singlet excited states; this produces 
an interesting structure for the specific heat at low temperature. The
absence of a singlet excitation gap also suggests that there may be a 
nonmagnetic long-range order, but the nature of this order
is not well-understood.

Classically, these systems often have an enormous ground state degeneracy
arising from local degrees of freedom which cost no energy; this leads to an
extensive entropy at zero temperature. Quantum mechanically, this
degeneracy is lifted, but one might still expect a remnant of the 
classical degeneracy in the form of a large number of low-energy excitations.
It is therefore useful to develop simple ways of understanding the
low-energy quantum excitations. In this paper, we will critically examine
one such way which is to consider the system as being made out of triangles 
which are weakly coupled to each other \cite{pert1,pert2,pert3}. We
first find the ground states of a single triangle assuming all the
three couplings to be antiferromagnetic with a strength $J_1$. We then use 
degenerate perturbation theory to study what happens when different triangles 
are coupled to each other with antiferromagnetic bonds of strength
$J_2$, where $J_2 \ll J_1$. Since experimental systems typically
have $J_2 = J_1$, we must finally study how accurate the perturbation 
theory is when the ratio $J_2 / J_1$ approaches $1$.

In Sec. II, we discuss the low-energy states of a single triangle of 
equal spins coupled antiferromagnetically to each other. If the spin at
each site is an integer, then there is a unique ground state for the 
triangle and the problem is not very interesting. But if the site spin $S$
is a half-odd-integer, then the ground state has a four-fold
degeneracy. This is because the ground state has total spin-$1/2$,
and there is an additional factor of two which can be interpreted
as the chirality. The chirality can be expressed as the eigenvalue of a
pseudospin operator which also has spin-$1/2$.
We then discuss the low-energy effective Hamiltonian (LEH) for 
a system in which each pair of triangles is coupled to each other by not 
more than one antiferromagnetic bond. To first order in $J_2 /J_1$,
the spins and pseudospins of two triangles get coupled to each other.
To second order, up to three triangles can get coupled to each other.
We have obtained the first order LEH (LEH1) for all values of $S$, and
the second order LEH (LEH2) for $S$ equal to $1/2$ and $3/2$.
We have not gone beyond second order because the LEHs rapidly
becomes more complicated and longer range as we go to higher orders. For
small systems, we numerically diagonalize the LEH1 and LEH2 
to obtain the energies of the ground state and the first excited state 
as functions of $J_2 / J_1$.

In Sec. III, we use both exact diagonalization for small systems and the 
density-matrix renormalization group method (DMRG) for larger systems to 
obtain the low-lying energies for the complete Hamiltonian. The DMRG 
is currently the most accurate numerical method known for obtaining the
low-energy properties of quantum systems in one-dimension \cite{whit1}. We
briefly describe the DMRG procedure for our system, and then compare the 
energies obtained from the LEHs and DMRG. Depending on the quantities of 
interest, we find that the results from the LEHs start deviating 
significantly from the DMRG values once $J_2 /J_1$ exceeds $0.2 ~-~ 0.4$. 
This is true even for the LEH2 which is significantly more 
accurate than the LEH1 for small values of $J_2 /J_1$. 

We also examine a simple system of five triangles
coupled to each other in such a way as to form a small fragment of the 
two-dimensional Kagome lattice; we find that the LEHs 
starts deviating from the exact results at even smaller values of $J_2 /J_1$. 
Thus the LEH approach becomes less accurate the more two-dimensional the
geometry is, i.e., the larger the coordination number is for each triangle. 

\section{Low-energy Hamiltonians for a chain of triangles}

We are interested in the system of spins shown in Fig. 1. The sites
are labeled as $(n,a)$, where $n$ labels the triangle and $a=1,2,3$ labels 
the three sites of a triangle as indicated. The spin at each site has the 
value $S$, and all the couplings are antiferromagnetic. The Hamiltonian is
\beq
{\hat H} ~=~ \sum_n ~\Bigl[~ J_1 ~(~ {\hat {\bf S}}_{n,1} \cdot {\hat {\bf 
S}}_{n,2} ~+~ {\hat {\bf S}}_{n,2} \cdot {\hat {\bf S}}_{n,3} ~+~ {\hat {\bf 
S}}_{n,3} \cdot {\hat {\bf S}}_{n,1} ~)~ +~ J_2 ~{\hat {\bf S}}_{n,2} \cdot {
\hat {\bf S}}_{n+1,3} ~]~.
\label{ham1}
\eeq
It is convenient to set $J_1 =1$ and consider only the parameter $J_2$.
 
Let us first examine a single triangle and drop the label $n$. The 
Hamiltonian ${\hat h}_{\Delta}$ is proportional to the square of the total 
spin operator, ignoring a shift in the zero of energy,
\beq
{\hat h}_{\Delta} ~=~ {\hat {\bf S}}_1 \cdot {\hat {\bf S}}_2 ~+~ {\hat {\bf 
S}}_2 \cdot {\bf S}_3 ~+~ {\hat {\bf S}}_3 \cdot {\hat {\bf S}}_1 ~=~ 
\frac{1}{2} ~ (~ {\hat {\bf S}}_1 ~+~ {\hat {\bf S}}_2 ~+~ {\hat {\bf S}}_3 ~
)^2 ~ -~ \frac{3}{2} ~S (S+1) ~.
\label{ham2}
\eeq
If $S$ is an integer ($1,2,...$), then the ground state of this triangle
is unique and is a singlet. Thus the low-energy sector of the entire system 
consists of only one state when different triangles are coupled to each 
other. The situation is much more interesting if $S$ is a half-odd-integer, 
i.e., $1/2,3/2,...~$. Then the ground state of each triangle is four-fold 
degenerate because it must have total spin $1/2$ and there are two ways of 
obtaining total spin-$1/2$ by combining three equal spins. For instance, 
sites $2$ and $3$ can first combine to give total spin $S_{23}$ equal to
either $S+1/2$ or $S-1/2$; $S_{23}$ can then combine with the first spin to 
produce a total spin-$1/2$. Since the total spin operator ${\hat {\bf S}} = 
{\hat {\bf S}}_1 + {\hat {\bf S}}_2 + {\hat {\bf S}}_3$ commutes with all the 
six permutations of the three spins, it is clear that the two ground states 
of the Hamiltonian in (\ref{ham2}) (with a given value of the total spin 
component $S^z = \pm 1/2$) must form a two-dimensional representation of the 
permutation group. This representation can be made explicit by introducing 
pseudospin-$1/2$ operators ${\hat {\bf \tau}}$ as follows. 

We will assume henceforth that all the site spins are identical and have the 
half-odd-integer value $S$. To obtain a total spin-$1/2$ for a triangle of 
three spins, the total spin of any two sites should be one of the integers 
$S \pm 1/2$. Under an exchange of the two spins $2$ and $3$, a state 
with their total spin $S_{23}$ transforms by the phase $(-1)^{S_{23}+1}$;
thus it is symmetric or antisymmetric depending on whether $S_{23}$ is odd or 
even. Let us define the pseudospin operator ${\hat \tau}^x$ such that the 
symmetric states have the eigenvalue of ${\hat \tau}^x = 1$, while the 
antisymmetric states have the eigenvalue of ${\hat \tau}^x = -1$.
Similarly, we can introduce an operator ${\hat \tau}^y$ such that states
which are symmetric (antisymmetric) with respect to exchange of spins $1$ and 
$2$ have eigenvalues of $(-{\hat \tau}^x + {\sqrt 3} {\hat \tau}^y )/2$ equal 
to $1$ ($-1$). For spins $1$ and $3$, the operator $(-{\hat \tau}^x - {\sqrt 
3} {\hat \tau}^y )/2$ will have the same property. We define the 
third pseudospin operator ${\hat \tau}^z = - (i/2) [{\hat \tau}^x , {\hat 
\tau}^y ]$. From these statements, it follows that within the space of the 
four ground states of a triangle, we have the operator identities
\bea
{\hat {\bf S}}_2 \cdot {\hat {\bf S}}_3 ~&=& ~A ~-~ \frac{(-1)^{S+1/2}}{4} ~
(2S+1) ~ {\hat \tau}^x ~, \nonumber \\
{\hat {\bf S}}_3 \cdot {\hat {\bf S}}_1 ~&=& ~A ~-~ \frac{(-1)^{S+1/2}}{4} ~
(2S+1) ~\Bigl[ ~- ~\frac{1}{2} {\hat \tau}^x ~-~ \frac{\sqrt 3}{2} {\hat 
\tau}^y ~\Bigr] ~, \nonumber \\
{\hat {\bf S}}_1 \cdot {\hat {\bf S}}_2 ~&=& ~A ~-~ \frac{(-1)^{S+1/2}}{4} ~
(2S+1) ~\Bigl[ ~ - ~\frac{1}{2} {\hat \tau}^x ~+~ \frac{\sqrt 3}{2} {\hat
\tau}^y ~\Bigr] ~, 
\label{iden}
\eea
where $A = -S^2 /2 - S/2 +1/8$. On taking the commutator of any two
of the equations in (\ref{iden}), we find that the three-spin chirality
operator is given by
\beq
{\hat {\bf S}}_1 \cdot {\hat {\bf S}}_2 \times {\hat {\bf S}}_3 ~=~ \frac{
\sqrt 3}{4} ~(~ S ~+~ \frac{1}{2} ~)^2 ~{\hat \tau}^z ~.
\label{chiral}
\eeq

Before proceeding further, we should mention that the pseudospin operators 
have been discussed earlier in Refs. 7 and 8 for the spin-$1/2$ 
case. In those papers, the two ground states (with a given 
value of $S^z$) are written in terms of a spin wave running around the 
triangle with momenta $\pm 2\pi /3$; these are called right and left moving 
respectively, and they are eigenstates of the operator ${\hat \tau}^z$. We 
have instead chosen to describe the states in terms of two of the spins 
forming total spin $S+1/2$ or $S-1/2$. The two descriptions are clearly
related to each other by an unitary transformation.

To continue, the Wigner-Eckart theorem says that the matrix elements of any of 
the site spin operators ${\hat {\bf S}}_a$ (where $a=1,2,3$) between the four 
ground states of a triangle must be proportional to the matrix elements of 
the total spin operator ${\hat {\bf s}}$. The proportionality 
factors must be independent of the spin component (i.e., $x$, $y$ or $z$), but 
they will involve the pseudospin operators. Let us introduce the operators
\beq
{\hat \tau}_a ~=~ \cos \frac{2\pi}{3} (1-a) ~{\hat \tau}^x ~+~ \sin 
\frac{2\pi}{3} (1-a) ~ {\hat \tau}^y ~,
\label{taua}
\eeq
where $a$ can take the values $1,2,3$. Using the spin wave functions, we can 
then show that the matrix elements of ${\hat {\bf S}}_a$ and ${\hat {\bf s}}$ 
are related as
\beq
\langle \sigma , \tau \vert {\hat {\bf S}}_a \vert \sigma^\prime , \tau^\prime 
\rangle ~=~ \frac{1}{3} ~\langle \sigma \vert {\hat {\bf s}} \vert 
\sigma^\prime \rangle ~ \Bigl[ ~\delta_{\tau , \tau^\prime} ~+~ (-1)^{S+1/2} ~
(2S+1) ~ \langle \tau \vert {\hat \tau}_a \vert \tau^\prime \rangle ~ \Bigr] ~,
\label{wigeck}
\eeq
where $\sigma , \sigma^\prime$ are the eigenstates of ${\hat \sigma}^z$ and
$\tau , \tau^\prime$ are the eigenstates of ${\hat \tau}^z$.

We can now derive the LEHs when two or more triangles are coupled together
as in Eq. (\ref{ham1}). We write that Hamiltonian as ${\hat H}= {\hat H}_0 + 
{\hat V}$, where ${\hat H}_0$ consists of the interactions within the 
triangles as in Eq. (\ref{ham2}), while $\hat V$ consists of the interactions 
between
triangles and is proportional to $J_2$. The LEH is a perturbative expansion in 
the parameter $J_2$. For $J_2 =0$, the low-energy sector for $N$ triangles
contains $4^N$ states, all of which have the same energy $E_0 = Ne_0$, where
\beq
e_0 ~=~ \frac{3}{8} ~-~ \frac{3}{2} ~S(S+1) ~.
\eeq
is the ground state energy of (\ref{ham2}). Following Ref. 9,
let us denote the different
low-energy states of the system as $p_i$ and the high-energy states
of the system as $q_{\alpha}$ (these are states in which at least one of the 
triangles is in a state with total spin $\ge 3/2$). The high-energy states 
have energies $E_{\alpha}$ according to the exactly solvable Hamiltonian 
${\hat H}_0$. Then the LEH1 is given by degenerate perturbation theory, 
\beq
{\hat H}_{eff}^{(1)} ~=~ \sum_{ij} ~\vert p_i \rangle ~\langle p_i \vert 
{\hat V} \vert p_j \rangle ~\langle p_j \vert ~.
\label{ham3}
\eeq
The LEH2 is given by
\beq
{\hat H}_{eff}^{(2)} ~=~ \sum_{ij} ~\sum_{\alpha} ~\vert p_i \rangle ~\frac{
\langle p_i \vert {\hat V} \vert q_{\alpha} \rangle ~\langle q_{\alpha} 
\vert {\hat V} \vert p_j \rangle}{E_0 ~-~ E_{\alpha}} ~\langle p_j \vert ~.
\label{ham4}
\eeq
The total effective Hamiltonian to this order is given by
\beq
{\hat H}_{LEH2} ~=~ E_0 ~+~ {\hat H}_{eff}^{(1)} ~+~ {\hat H}_{eff}^{(2)} ~.
\eeq

The LEH1 can now be directly read off from Eq. (\ref{wigeck}) after adding the 
triangle label $n$; thus the total spin operator of triangle $n$ is denoted 
as ${\hat {\bf s}}_n$ and the pseudospin operator as ${\hat {\bf \tau}}_n$. 
In general, if site number $a$ of triangle $l$ is connected to site number $b$ 
of triangle $m$ by a bond of strength $J_2$ (see Fig. 2), the contribution
of that bond to the LEH1 is given by 
\beq
{\hat h}_{eff}^{(1)} ~=~ \frac{J_2}{9} ~{\hat {\bf s}}_l \cdot {\hat {\bf 
s}}_m ~ \Bigl[ ~1 ~+~ (-1)^{S+1/2} ~ (2S+1) ~ {\hat \tau}_l^a ~\Bigl] ~
\Bigl[ ~1 ~+~ (-1)^{S+1/2} ~ (2S+1) ~{\hat \tau}_m^b ~ \Bigl] ~.
\label{ham5}
\eeq
For the particular form of couplings shown in Fig. 1, the total LEH1 takes the
form
\beq
{\hat H}_{LEH1} ~= - ~ Ne_0 + \frac{J_2}{9} ~\sum_n ~{\hat {\bf s}}_n 
\cdot {\hat {\bf s}}_{n+1} ~\Bigl[ 1 ~+~ (-1)^{S+1/2} ~ (2S+1) ~{\hat 
\tau}_n^2 ~\Bigl] ~\Bigl[ 1 ~+~ (-1)^{S+1/2} ~(2S+1) ~{\hat \tau}_{n+1}^3 ~ 
\Bigl] .
\label{ham6}
\eeq

The derivation of the LEH2 in (\ref{ham4}) requires a much 
longer calculation since we have to first compute matrix elements of the
form $\langle q_{\alpha} \vert {\hat {\bf S}}_a \vert p_i \rangle$ for all the
states within each triangle, and we then have to take products of these
to obtain the matrix elements of ${\hat {\bf s}}_{n,2} \cdot {\hat {\bf 
s}}_{n+1,3}$. The second-order terms in the LEH can arise from either
(i) a single bond connecting site $a$ of triangle $l$ to site $b$ of triangle
$m$, or (ii) a bond connecting site $a$ of triangle $l$ to site $b$ of
triangle $m$ and another bond connecting site $c$ of triangle $m$ to site
$d$ of triangle $n$, where $a,b,c,d$ take values from $1,2,3$ (here $b$ may or
may not be equal to $c$). The notation for these two types is shown in Fig. 2.

Let us first consider the simplest situation in which all the site spins are 
equal to $S=1/2$. The contribution of type (i) to the LEH2 is then given by
\beq
{\hat h}_{eff,(i)}^{(2)} ~=~ - \frac{J_2^2}{54} ~\Bigl[ ~(1- {\hat 
\tau}_{l,a} { \hat \tau}_{m,b} )~(3 +4 {\hat {\bf s}}_l \cdot {\hat {\bf 
s}}_m ) ~+~ (1+ { \hat \tau}_{l,a} )(1+ {\hat \tau}_{m,b} )~\Bigr] ~,
\label{ham7}
\eeq
where we have used the notation of Eq. (\ref{taua}). The contribution of 
type (ii) is
\bea
{\hat h}_{eff,(ii)}^{(2)} ~=~ - \frac{4J_2^2}{243} ~(1- 2 {\hat \tau}_{l,a} 
)(1- 2 {\hat \tau}_{n,d} ) ~\Bigl[ ~& & \Bigl( ~\cos \frac{2\pi}{3} (b-c) ~+~ 
{\hat \tau}_{m,-b-c} ~\Bigr) ~{\hat {\bf s}}_l \cdot {\hat {\bf s}}_n 
\nonumber \\
& & +~ \sin \frac{2\pi}{3} (b-c) ~{\hat \tau}_m^z ~{\hat {\bf s}}_l \cdot 
{\hat {\bf s}}_m \times {\hat {\bf s}}_n ~\Bigr] ~.
\label{ham8}
\eea
Putting all this together for the system in Fig. 1, we find that for
$S=1/2$, the total LEH2 is given by
\bea
{\hat H}_{LEH2} ~=~ & & - \frac{3N}{4} ~+~ \frac{J_2}{9} ~\sum_n ~{\hat 
{\bf s}}_n \cdot {\hat {\bf s}}_{n+1} ~( 1 - 2 {\hat \tau}_{n,2} )~( 1 - 2 
{\hat \tau}_{n+1,3} ) \nonumber \\
& & - \frac{J_2^2}{54} ~\sum_n ~\Bigl[ ~(1- {\hat \tau}_{n,2} {\hat 
\tau}_{n+1,3} )~(3 +4 {\hat {\bf s}}_n \cdot {\hat {\bf s}}_{n+1} ) ~+~ (1+ 
{\hat \tau}_{n,2} ) ~(1+ {\hat \tau}_{n+1,3} )~\Bigr] \nonumber \\
& & - \frac{4J_2^2}{243} ~\sum_n ~(1- 2 {\hat \tau}_{n,2} ) ~(1- 2 {\hat 
\tau}_{n+1,3} )~ \Bigl[ ( - \frac{1}{2} + {\hat \tau}_{n+1,1} ) ~{\hat {\bf 
s}}_n \cdot {\hat {\bf s}}_{n+2} - \frac{\sqrt 3}{2} ~{\hat \tau}_{n+1}^z ~
{\hat {\bf s}}_n \cdot {\hat {\bf s}}_{n+1} \times {\hat {\bf s}}_{n+2} 
\Bigr]  ~. \nonumber \\
& &
\label{ham9}
\eea

If the site spins are equal to $S=3/2$, the contribution of type (i) 
to the LEH2 is given by
\bea
{\hat h}_{eff,(i)}^{(2)} ~=~ - \frac{J_2^2}{27} ~\Bigl[ ~ & & 56 ~+~ 
42 {\hat {\bf s}}_l \cdot {\hat {\bf s}}_m ~+~ ({\hat \tau}_{l,a} ~+~ {\hat 
\tau}_{m,b} ) ~(-1 ~+~ 8 {\hat {\bf s}}_l \cdot {\hat {\bf s}}_m ) 
\nonumber \\
& & - {\hat \tau}_{l,a} {\hat \tau}_{m,b} (4 ~+~ 8 {\hat {\bf s}}_l \cdot 
{\hat {\bf s}}_m ) ~.
\label{ham10}
\eea
The contribution of type (ii) is
\bea
{\hat h}_{eff,(ii)}^{(2)} ~=~ - \frac{4J_2^2}{243} ~(1+4 {\hat \tau}_{l,a} 
)(1+4 {\hat \tau}_{n,d} ) ~ \Bigl[ ~& & \Bigl( ~7 ~\cos \frac{2\pi}{3} (b-
c) ~-~ 2~ { \hat \tau}_{m,-b-c} ~ \Bigr) ~{\hat {\bf s}}_l \cdot {\hat {\bf 
s}}_n \nonumber \\
& & +~ 4 ~\sin \frac{2\pi}{3} (b-c) ~{\hat \tau}_m^z ~{\hat {\bf s}}_l \cdot 
{\hat {\bf s}}_m \times {\hat {\bf s}}_n ~\Bigr] ~.
\label{ham12}
\eea
For the system in Fig. 1, the total LEH2 for $S=3/2$ is therefore
\bea
{\hat H}_{LEH2} ~=~ & & - \frac{21N}{4} ~+~ \frac{J_2}{9} ~\sum_n ~{\hat 
{\bf s}}_n \cdot {\hat {\bf s}}_{n+1} ~( 1 + 4 {\hat \tau_{n,2}} )~( 1 + 4 
{\hat \tau}_{n+1,3} ) \nonumber \\
& & - \frac{J_2^2}{27} ~\sum_n ~\Bigl[ 56 + 42 {\hat {\bf s}}_n \cdot {\hat 
{\bf s}}_{n+1} + ({\hat \tau}_{n,2} + {\hat \tau}_{n+1,3}) (-1 + 8 {\hat {\bf 
s}}_n \cdot {\hat {\bf s}}_{n+1}) - {\hat \tau}_{n,2} {\hat \tau}_{n+1,3} (4 
+8 {\hat {\bf s}}_n \cdot {\hat {\bf s}}_{n+1} )\Bigr] 
\nonumber \\
& & - \frac{4J_2^2}{243} ~\sum_n ~(1+ 4 {\hat \tau}_{n,2} ) ~(1+ 4 {\hat 
\tau}_{n+1,3} )~ \Bigl[ ( - \frac{7}{2} + {\hat \tau}_{n+1,1} ) ~{\hat {\bf 
s}}_n \cdot {\hat {\bf s}}_{n+2} - 2 {\sqrt 3} ~{\hat \tau}_{n+1}^z ~{\hat 
{\bf s}}_n \cdot {\hat {\bf s}}_{n+1} \times {\hat {\bf s}}_{n+2} \Bigr] . 
\nonumber \\
& &
\label{ham13}
\eea

We have carried out exact diagonalization studies of the LEH1 and LEH2 for 
systems up to $10$ triangles for both the spin-$1/2$ and spin-$3/2$ cases.

\section{Density-matrix renormalization group study of the chain of triangles}

We have numerically studied the system described by Eq. (\ref{ham1})
using exact diagonalization for small systems and the DMRG method for larger
systems. The number of triangles is $N$ and the number of sites is $3N$.

For small systems, we have performed exact diagonalization with periodic 
boundary conditions. For larger systems, we have done 
DMRG calculations (using the infinite system algorithm \cite{whit1}) with 
open boundary conditions. For exact diagonalization, we have gone up to $30$ 
sites ($10$ triangles). With DMRG, we have gone up to $50$ triangles and
in some cases up to $100$ triangles, after checking that the DMRG 
and exact results match for $10$ triangles for the spin-$1/2$ system.
The number of dominant density matrix eigenstates, corresponding to
the $m$ largest eigenvalues of the density matrix, that we retained 
at each DMRG iteration was $64$. In fact, we varied the value of $m$ 
from $64$ to $100$ and found that $m=64$ gives satisfactory results in 
terms of agreement with exact diagonalization for small systems and good 
numerical convergence for large systems. The system is grown by adding two 
new triangles at each iteration; we found that this gives more accurate 
results than either adding two new sites at each iteration (in which case 
we would have obtained the triangle structure only after every third 
iteration) or adding one triangle at each iteration. 

In Fig. 3, we show the DMRG values of the ground state energy per site 
versus $1/N$ for a few illustrative values of $J_2$ (in units of $J_1$) 
for the case in which each site has spin-$1/2$. After extrapolating
to the thermodynamics limit $N \rightarrow \infty$, we show the ground state
energy per site as a function of $J_2$ in Fig. 4. In that figure, we
also show the numerical results obtained by exact diagonalization of the
LEH1 and LEH2 respectively. As expected, the LEH2 is more accurate than 
the LEH1 up to a larger value of $J_2$. However, even the LEH2 becomes 
fairly inaccurate beyond about $J_2 = 0.4$. 

A more detailed comparison of the LEH1 and LEH2 with the exact results for 
chains with up to $10$ triangles is given in Table 1. We see that the LEH2 
agrees better with the exact results than the LEH1, although the agreement 
becomes rather poor for large $J_2$. We also see 
that the accuracy of the LEHs is poorer for the gap than it is
for the ground state energy. For instance, the LEH2 results for the
gap become relatively inaccurate even for $J_2 = 0.2$.

Fig. 5 shows the low-lying triplet and singlet gaps above the ground state for 
the same system at different values of $J_2$. The figure compares the results 
obtained from the LEH2 with those obtained after extrapolating the gaps from 
exact diagonalization of the full Hamiltonian for different values of $N$. 
Both the triplet (Fig. 5 (a)) and singlet (Fig. 5 (b)) gaps deviate 
considerably from the "exact" results beyond $J_2 \sim 0.4$. In fact, we see 
from Fig. 5 (a) that the ground state of the LEH2 switches from a singlet to 
a triplet for $J_2 > 0.7$.

In Fig. 6, we show a larger number of $S=0$ and $S=1$ states which have very 
small gaps above the ground state. It is possible that some of these actually 
become degenerate with the ground state in the thermodynamic limit. The lowest 
spin sector with an appreciable gap (which is likely to remain non-zero in the 
thermodynamic limit) seems to be $S=2$. However it is possible that there
are many more singlet and triplet states lying below the lowest $S=2$ state
than we have shown in the figure; it is very difficult to find more than
a few low-lying states in each spin sector using the DMRG method. For the
same reason, we cannot rule out the possibility of a finite gap to a 
higher singlet or a higher triplet (than what we have targetted)
lying below the quintet. Fig. 7 shows an even larger number of 
low-lying states for the case $J_2 = 1.0$. The lowest three 
states with $S^z =0$ (including the ground state), the lowest six states with 
$S^z =0$ and the lowest state with $S^z =2$ are shown there. To
summarize, we have found an unexpectedly large number of low-lying (and
possibly gapless) singlet and triplet excitations in this system. We will 
provide an explanation for some of these states below. We note that
the low-energy spectrum has a resemblance to that found in the 
Kagome lattice in two dimensions \cite{lech}. However there is an important 
difference between the two cases; the gapless band of excitations in 
the Kagome problem consists only of singlets, and the first gap
is to a triplet state.

In Fig. 8 (a), we show the DMRG values of the bond order 
for the middle two bonds, namely, the ground state expectation values
$\langle {\hat {\bf s}}_{p,2} \cdot {\hat {\bf s}}_{p+1,3} \rangle$ for 
$p = N/2 -1$ and $N$, as a function of $N$ at $J_2 =1.0$ for the spin-$1/2$ 
system. In Fig. 8 (b), we plot the bond order alternation (or spontaneous 
dimerization), defined to be the magnitude of the difference of the two 
neighbouring bond orders in Fig. 8 (a) extrapolated to the 
limit $N \rightarrow \infty$, as a function of $J_2$. We observe 
that the alternation is quite large for small values of $J_2$ and that it
remains non-zero even at a large value of $J_2 =J_1$. Following Mila 
\cite{pert2}, we can qualitatively understand the dimerization occurring 
at small values of $J_2$ as follows. Since the spin and pseudospin degrees of 
freedom appear very asymmetrically in the LEHs, we perform a mean-field
decoupling of these two. We assume that the pseudospin variables take some 
fixed values and find the quantum ground state of the spin variables in that 
fixed background. For two triangles coupled together by the LEH1 
in Eq. (\ref{ham5}), we see that the lowest energy is attained if the 
pseudospin variables satisfy $\tau_{l,a} = \tau_{m,b} = -1$ (for the
case $S=1/2$), and the effective spin-$1/2$ of the two triangles then form a
singlet. In the same way, the mean-field ground state of the chain is
given by the dimerized configuration in which $\tau_{2n,2} = \tau_{2n+1,3} =
-1$ and the effective spin-$1/2$ of triangles $2n$ and $2n+1$ form a 
singlet. Such a dimerized ground state can also be obtained by translating 
the above state by one triangle; it is therefore two-fold degenerate as in a
spin-Peierls system.

It is instructive to contrast this system with the antiferromagnetic 
spin-$1/2$ chain with a nearest-neighbor coupling $J_1$ and a 
next-nearest-neighbor coupling $J_2$ \cite{whit2}. For $J_2 /J_1 > 0.241...$,
this is known to spontaneously dimerize, and there is also a finite gap to 
excitations in the bulk and a finite correlation length $\xi$ in the 
thermodynamic limit. In the dimerized phase, an open chain with an even 
number of sites has five ground states corresponding to two singlets and 
one triplet. One singlet and one triplet arise from the free 
spin-$1/2$ degrees of freedom which reside at the two ends of the chain. 
The finite correlation length $\xi$ means that the splitting between 
these states vanishes exponentially with the size of the system.

Similarly, for our system with a chain of an even number of triangles, we
may expect at least some of the low-energy states to arise from the effective
spin-$1/2$ degrees of freedom residing on the two end triangles.
Due to the presence of the pseudospin-$1/2$ degrees of freedom in each
triangle, we expect that there will be a total of $4^2 =16$ low-energy 
states forming four triplets and four singlets.
The number of low-energy singlets and triplets that we actually find is more
than this; this implies that there are some additional low-energy degrees
of freedom (probably associated with the bulk) which we do not yet understand.

In Table 2, we show the ground state energy per site versus $J_2$ for the case 
in which each site has spin-$3/2$. The numerical results obtained by exact 
diagonalization of the LEH1 and LEH2 are also shown. A comparison with Table 1
for the spin-$1/2$ case shows that the LEH2 starts deviating from the exact 
results for smaller values of $J_2$ as the site spin is increased.

Finally, in Fig. 9, we show the ground state energy per site versus $J_2$ for 
a group of five triangles forming a sub-system of a two-dimensional Kagome 
lattice. The reason for studying this system is to test how well the LEHs do
as the coordination number of the triangles is increased from two, thereby
making the system more two-dimensional. We find that the LEH2 fails even
faster with increasing $J_2$ than it does for the chains studied above.

\section{Summary and Outlook}

We have studied a chain of coupled triangles of half-odd-integer spins using 
both the DMRG method and the LEH approaches. We find that the LEH approach is
accurate for the ground state energy only for small values of the ratio of the 
coupling between triangles and the coupling within each triangle; the accuracy
for the low-energy gaps is less than that for the ground state energy. The 
range of accuracy of the LEHs is also smaller for larger values of the site 
spin as well as for a larger number of neighbors coupled to each triangle.
We therefore conclude that the LEH approaches may not be very reliable for 
the low-energy properties of the currently existing experimental systems in 
which all couplings are of the same order and the geometry is two-dimensional
\cite{pert2}.

\vskip .7 true cm
\noindent {\bf Acknowledgments}

We thank Kunj Tandon and Swapan Pati for useful discussions. SR thanks the
Department of Science and Technology, Govt. of India for financial support 
under project SP/S1/H-07/96.

\newpage

\begin{center}
\begin{tabular}{c c}
\hspace{1.8cm}{\bf{Ground State}}\hspace{3.0cm} {\bf{First Excited State}} \\
\end{tabular}
\begin{tabular}{|c|c|c|c|c||c|c|c|}
\hline ${\bf{J_{2}}}$ & {\bf{N}} & {\bf{LEH1}} & {\bf{LEH2}} &
{\bf{Exact }} & {\bf{LEH1}} & {\bf{LEH2}} &
{\bf{Exact }} \\ \hline 
&\hspace{0.1cm} 4 \hspace{0.1cm}&\hspace{0.2cm}-3.15843\hspace{0.2cm} &
\hspace{0.2cm} -3.16139\hspace{0.2cm} &\hspace{0.2cm} -3.16135\hspace{0.2cm} &
\hspace{0.2cm}-3.14912\hspace{0.2cm} &\hspace{0.2cm}-3.15131\hspace{0.2cm} &
\hspace{0.2cm}-3.15130\hspace{0.2cm} \\ \cline{2-8} 
0.1 &\hspace{0.1cm} 6\hspace{0.1cm} &\hspace{0.2cm} -4.73740\hspace{0.2cm} &
\hspace{0.2cm} -4.73772\hspace{0.2cm} &\hspace{0.2cm} -4.73768\hspace{0.2cm} &
\hspace{0.2cm} -4.72959\hspace{0.2cm} &\hspace{0.2cm} -4.73334\hspace{0.2cm} &
\hspace{0.2cm} -4.73331\hspace{0.2cm} \\ \cline{2-8}
&\hspace{0.1cm} 8\hspace{0.1cm} &\hspace{0.2cm} -6.30976\hspace{0.2cm} &
\hspace{0.2cm} -6.31541\hspace{0.2cm} &\hspace{0.2cm} -6.31536\hspace{0.2cm} &
\hspace{0.2cm} -6.30795\hspace{0.2cm} &\hspace{0.2cm} -6.31319\hspace{0.2cm} &
\hspace{0.2cm} -6.31315\hspace{0.2cm} \\ \hline \hline
&\hspace{0.1cm} 4\hspace{0.1cm} &\hspace{0.2cm} -3.31686\hspace{0.2cm}  &
\hspace{0.2cm} -3.32892\hspace{0.2cm} &\hspace{0.2cm} -3.32864\hspace{0.2cm} &
\hspace{0.2cm} -3.29825\hspace{0.2cm} &\hspace{0.2cm} -3.30728\hspace{0.2cm} &
\hspace{0.2cm} -3.30713\hspace{0.2cm} \\ \cline{2-8}
0.2 &\hspace{0.1cm} 6\hspace{0.1cm} &\hspace{0.2cm} -4.96681\hspace{0.2cm} &
\hspace{0.2cm} -4.98455\hspace{0.2cm} &\hspace{0.2cm} -4.98417\hspace{0.2cm} &
\hspace{0.2cm} -4.95918\hspace{0.2cm} &\hspace{0.2cm} -4.97455\hspace{0.2cm} &
\hspace{0.2cm} -4.97427\hspace{0.2cm} \\ \cline{2-8}
&\hspace{0.1cm} 8\hspace{0.1cm} &\hspace{0.2cm} -6.61952\hspace{0.2cm} &
\hspace{0.2cm} -6.64278\hspace{0.2cm} &\hspace{0.2cm} -6.64229\hspace{0.2cm} &
\hspace{0.2cm} -6.61590\hspace{0.2cm} &\hspace{0.2cm} -6.63740\hspace{0.2cm} &
\hspace{0.2cm} -6.63699\hspace{0.2cm} \\ \hline \hline
&\hspace{0.1cm} 4\hspace{0.1cm} &\hspace{0.2cm} -3.79215\hspace{0.2cm} &
\hspace{0.2cm} -3.87213\hspace{0.2cm} &\hspace{0.2cm} -3.86669\hspace{0.2cm} &
\hspace{0.2cm} -3.74562\hspace{0.2cm} &\hspace{0.2cm} -3.80782\hspace{0.2cm} &
\hspace{0.2cm} -3.80463\hspace{0.2cm} \\ \cline{2-8}
0.5 &\hspace{0.1cm} 6\hspace{0.1cm} &\hspace{0.2cm} -5.66702\hspace{0.2cm} &
\hspace{0.2cm} -5.78669\hspace{0.2cm} &\hspace{0.2cm} -5.77885\hspace{0.2cm} &
\hspace{0.2cm} -5.64794\hspace{0.2cm} &\hspace{0.2cm} -5.75145\hspace{0.2cm} &
\hspace{0.2cm} -5.74604\hspace{0.2cm} \\ \cline{2-8}
&\hspace{0.1cm} 8\hspace{0.1cm} &\hspace{0.2cm} -7.54881\hspace{0.2cm} &
\hspace{0.2cm} -7.70679\hspace{0.2cm} &\hspace{0.2cm} -7.69665\hspace{0.2cm} &
\hspace{0.2cm} -7.53975\hspace{0.2cm} &\hspace{0.2cm} -7.68454\hspace{0.2cm} &
\hspace{0.2cm} -7.67639\hspace{0.2cm} \\ \hline \hline
&\hspace{0.1cm} 4\hspace{0.1cm} &\hspace{0.2cm} -4.58430\hspace{0.2cm} &
\hspace{0.2cm} -4.93506\hspace{0.2cm} &\hspace{0.2cm} -4.88000\hspace{0.2cm} &
\hspace{0.2cm} -4.49125\hspace{0.2cm} & \hspace{0.2cm}-4.78367 &
\hspace{0.2cm} -4.74638\hspace{0.2cm} \\ \cline{2-8}
1.0 &\hspace{0.1cm} 6\hspace{0.1cm} &\hspace{0.2cm} -6.83404\hspace{0.2cm} &
\hspace{0.2cm} -7.02622\hspace{0.2cm} &\hspace{0.2cm} -7.28614\hspace{0.2cm} &
\hspace{0.2cm} -6.79589\hspace{0.2cm} &\hspace{0.2cm} -7.26277\hspace{0.2cm} &
\hspace{0.2cm} -7.20489\hspace{0.2cm} \\ \cline{2-8}
&\hspace{0.1cm} 8\hspace{0.1cm} &\hspace{0.2cm} -9.09761\hspace{0.2cm} &
\hspace{0.2cm} -9.81245\hspace{0.2cm} &\hspace{0.2cm} -9.70079\hspace{0.2cm} &
\hspace{0.2cm} -9.07950\hspace{0.2cm} &\hspace{0.2cm} -9.73152\hspace{0.2cm} &
\hspace{0.2cm} -9.64368\hspace{0.2cm} \\ \hline
\end{tabular}
\end{center}
\vskip 1.2 true cm

\noindent Table 1. Ground state and first excited state energies in units of
$J_1$ obtained for the spin-$1/2$ system using exact diagonalization of the 
full Hamiltonian and the LEH1 and LEH2.

\vskip 2.8 true cm

\begin{center}
\hspace{-3cm}
\begin{tabular}{c c}
\hspace{4.0cm}{\bf{N=3}} & \hspace{5.0cm}{\bf{N=4}} \\ 
\end{tabular}
\begin{tabular}{|c|c|c|c||c|c|c|}
\hline ${\bf{J_{2}}}$ &  {\bf{LEH1}} & {\bf{LEH2}} &
{\bf{Exact }} &  {\bf{LEH1}} & {\bf{LEH2}} &
{\bf{Exact }}  \\ \hline
\hspace{0.2cm}0.1\hspace{0.2cm} &\hspace{0.2cm}  -16.0462\hspace{0.2cm} &
\hspace{0.2cm} -16.1044\hspace{0.2cm} &\hspace{0.2cm} -16.1071\hspace{0.2cm} &
\hspace{0.2cm} -21.4577\hspace{0.2cm} &\hspace{0.2cm}  -21.5211
\hspace{0.2cm} & \hspace{0.2cm} -21.5224\hspace{0.2cm} \\ \hline
\hspace{0.2cm}0.2\hspace{0.2cm} &\hspace{0.2cm}  -16.3424\hspace{0.2cm} &
\hspace{0.2cm} -16.5793\hspace{0.2cm} &\hspace{0.2cm} -16.5692 
\hspace{0.2cm}  &\hspace{0.2cm} -21.9154\hspace{0.2cm} &
\hspace{0.2cm} -22.1775\hspace{0.2cm} &\hspace{0.2cm} -22.1611
\hspace{0.2cm} \\ \hline
\hspace{0.2cm}0.3\hspace{0.2cm}  & \hspace{0.2cm} -16.6386\hspace{0.2cm} &
\hspace{0.2cm} -17.1799\hspace{0.2cm} &\hspace{0.2cm} -17.1132\hspace{0.2cm} &
\hspace{0.2cm} -22.3731\hspace{0.2cm} &\hspace{0.2cm} -22.9829\hspace{0.2cm} &
\hspace{0.2cm} -22.8961\hspace{0.2cm} \\ \hline
\hspace{0.2cm}0.4 \hspace{0.2cm}  & \hspace{0.2cm} -16.9348\hspace{0.2cm} &
\hspace{0.2cm} -17.9107\hspace{0.2cm} &\hspace{0.2cm} -17.7212\hspace{0.2cm} &
\hspace{0.2cm} -22.8308\hspace{0.2cm} &\hspace{0.2cm} -23.9508\hspace{0.2cm} &
\hspace{0.2cm} -23.7107\hspace{0.2cm} \\ \hline
\hspace{0.2cm}0.5\hspace{0.2cm} & \hspace{0.2cm} -17.2309\hspace{0.2cm} &
\hspace{0.2cm} -18.7747\hspace{0.2cm} &\hspace{0.2cm} -18.3807\hspace{0.2cm} & 
\hspace{0.2cm} -23.2884\hspace{0.2cm} &\hspace{0.2cm} -25.0933\hspace{0.2cm} &
\hspace{0.2cm} -24.5916\hspace{0.2cm} \\ \hline
\hspace{0.2cm}0.6\hspace{0.2cm} & \hspace{0.2cm}  -17.5271\hspace{0.2cm} &
\hspace{0.2cm} -19.7746\hspace{0.2cm} &\hspace{0.2cm} -19.0834\hspace{0.2cm} &
\hspace{0.2cm} -23.7461\hspace{0.2cm} &\hspace{0.2cm} -26.4200\hspace{0.2cm} & 
\hspace{0.2cm}-25.5294\hspace{0.2cm} \\ \hline
\hspace{0.2cm}0.7\hspace{0.2cm} & \hspace{0.2cm} -17.8233\hspace{0.2cm} &
\hspace{0.2cm} -20.9118\hspace{0.2cm} & \hspace{0.2cm} -19.8233 
\hspace{0.2cm} & \hspace{0.2cm} -24.2038\hspace{0.2cm} &
\hspace{0.2cm} -27.9539\hspace{0.2cm} &\hspace{0.2cm} -26.5170
\hspace{0.2cm} \\ \hline
\hspace{0.2cm}0.8\hspace{0.2cm} & \hspace{0.2cm} -18.1195\hspace{0.2cm} &
\hspace{0.2cm} -22.1878\hspace{0.2cm} &\hspace{0.2cm} -20.5966 
\hspace{0.2cm} & \hspace{0.2cm} -24.6615\hspace{0.2cm} &
\hspace{0.2cm} -29.7650\hspace{0.2cm} &\hspace{0.2cm} -27.5488
\hspace{0.2cm} \\ \hline
\hspace{0.2cm}0.9 \hspace{0.2cm}  & \hspace{0.2cm} -18.4157\hspace{0.2cm} &
\hspace{0.2cm} -23.6032\hspace{0.2cm} &\hspace{0.2cm} -21.4006
\hspace{0.2cm} & \hspace{0.2cm} -25.1192\hspace{0.2cm} &
\hspace{0.2cm} -31.8070\hspace{0.2cm} &\hspace{0.2cm} -28.6206
\hspace{0.2cm} \\ \hline
\hspace{0.2cm}1.0\hspace{0.2cm} &\hspace{0.2cm} -18.7119\hspace{0.2cm} &
\hspace{0.2cm} -25.1586\hspace{0.2cm} &\hspace{0.2cm} -22.2325\hspace{0.2cm} &
\hspace{0.2cm} -25.5769\hspace{0.2cm} & \hspace{0.2cm} -34.0823
\hspace{0.2cm} & \hspace{0.2cm} -29.7285\hspace{0.2cm} \\ \hline
\end{tabular}
\end{center}
\vskip 1.2 true cm

\noindent Table 2. Ground state energies obtained for the spin-$3/2$ system 
using exact diagonalization of the full Hamiltonian and the LEH1 and LEH2 for 
$3$ and $4$ triangles.

\newpage

\noindent {\bf Figure Captions}
\vskip .5 true cm

\noindent {1.} The chain of triangles showing the labeling of the vertices
of each triangle and the antiferromagnetic couplings $J_1 = 1$ and $J_2$.

\noindent{2.} The triangle and site labels used in deriving the LEHs. 

\noindent
{3.} The ground state energy/site in units of $J_1$ vs $1/N$, for 
a few values of $J_2$ with spin-$1/2$ at each site. 

\noindent
{4.} The ground state energy/site in units of $J_1$, for values of $J_2 = 
0.1$ to $1.0$. The three sets of points show the results using DMRG for the
full Hamiltonian and the results from the LEH1 and LEH2.

\noindent
{5.} The energy gaps from the ground state in units of $J_1$, for values of 
$J_2 = 0.1$ to $1.0$ for the spin-$1/2$ system. (a) shows the triplet gap and 
(b) shows the singlet gap. The two sets of points show the results obtained
from the LEH2 and DMRG.

\noindent
{6.} Some low-lying energy gaps for $J_2 = 0.0$ to $1.0$ for the spin-$1/2$ 
system obtained by DMRG. The two lowest $S^z =0$ states (including the 
ground state at the bottom of the figure), the two lowest $S^z =1$ states 
and the lowest $S^z =2$ state are shown.

\noindent
{7.} A larger number of low-lying energy gaps for $J_2 = 1.0$ for the 
spin-$1/2$ system obtained by DMRG. The two lowest $S^z =0$ states (including 
the ground state at the bottom), the six lowest $S^z =1$ states and the 
lowest $S^z =2$ state are shown.

\noindent
{8.} The spin bond orders for the spin-$1/2$ system obtained by DMRG. 
(a) shows the bond orders for two neighbouring bonds in the middle
of the chain as a function of $N$ for $J_2 =1.0$. 
(b) shows the difference of the two bond orders (called the bond alternation 
or dimerization) for various values of $J_2$.

\noindent
{9.} The ground state energy in units of $J_1 = 1$, for values of $J_2 =
0.1$ to $1.0$ for spin-$1/2$ at each site of a system of five triangles
as shown in the inset. The two sets of points show the results from the
exact diagonalization of the LEH2 and of the full Hamiltonian.

\newpage

\begin{figure}[ht]
\vspace*{3cm}
\begin{center}
\epsfig{figure=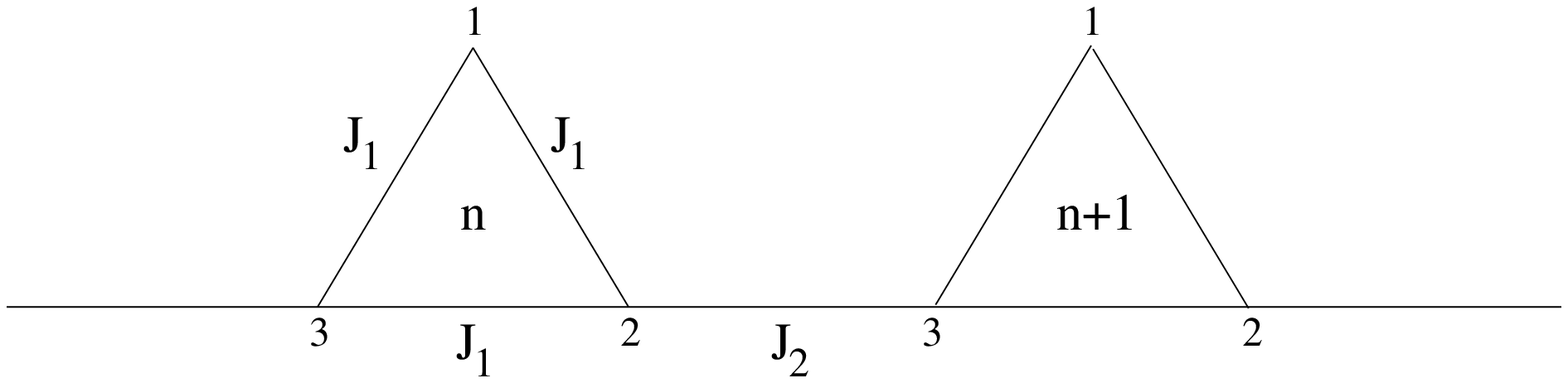,width=15cm}
\end{center}
\vspace*{1cm}
\centerline{Fig. 1}
\label{fig1}
\end{figure}

\vspace*{5cm}
\begin{figure}[hp]
\begin{center}
\epsfig{figure=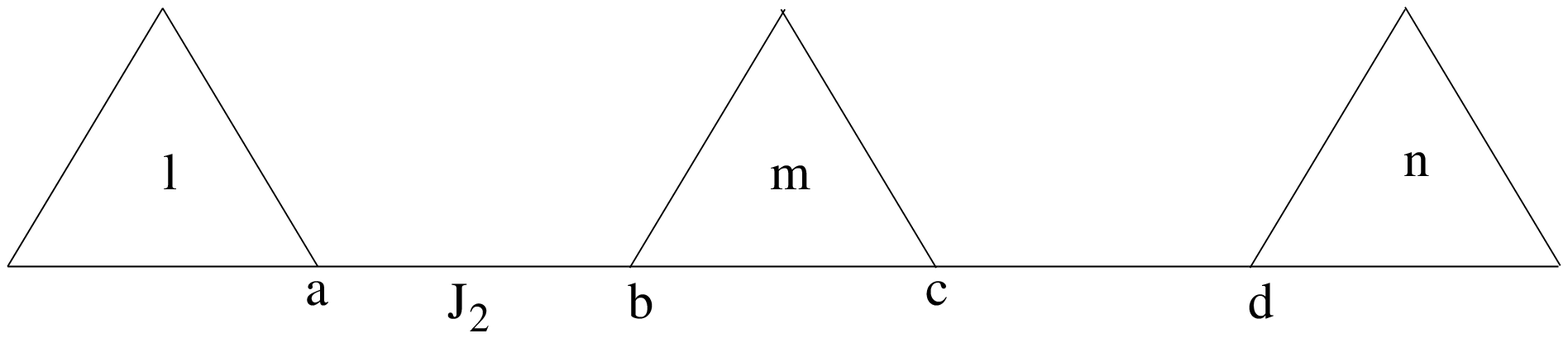,width=15cm}
\end{center}
\vspace*{1cm}
\centerline{Fig. 2}
\label{fig2}
\end{figure}

\begin{figure}[hp]
\begin{center}
\epsfig{figure=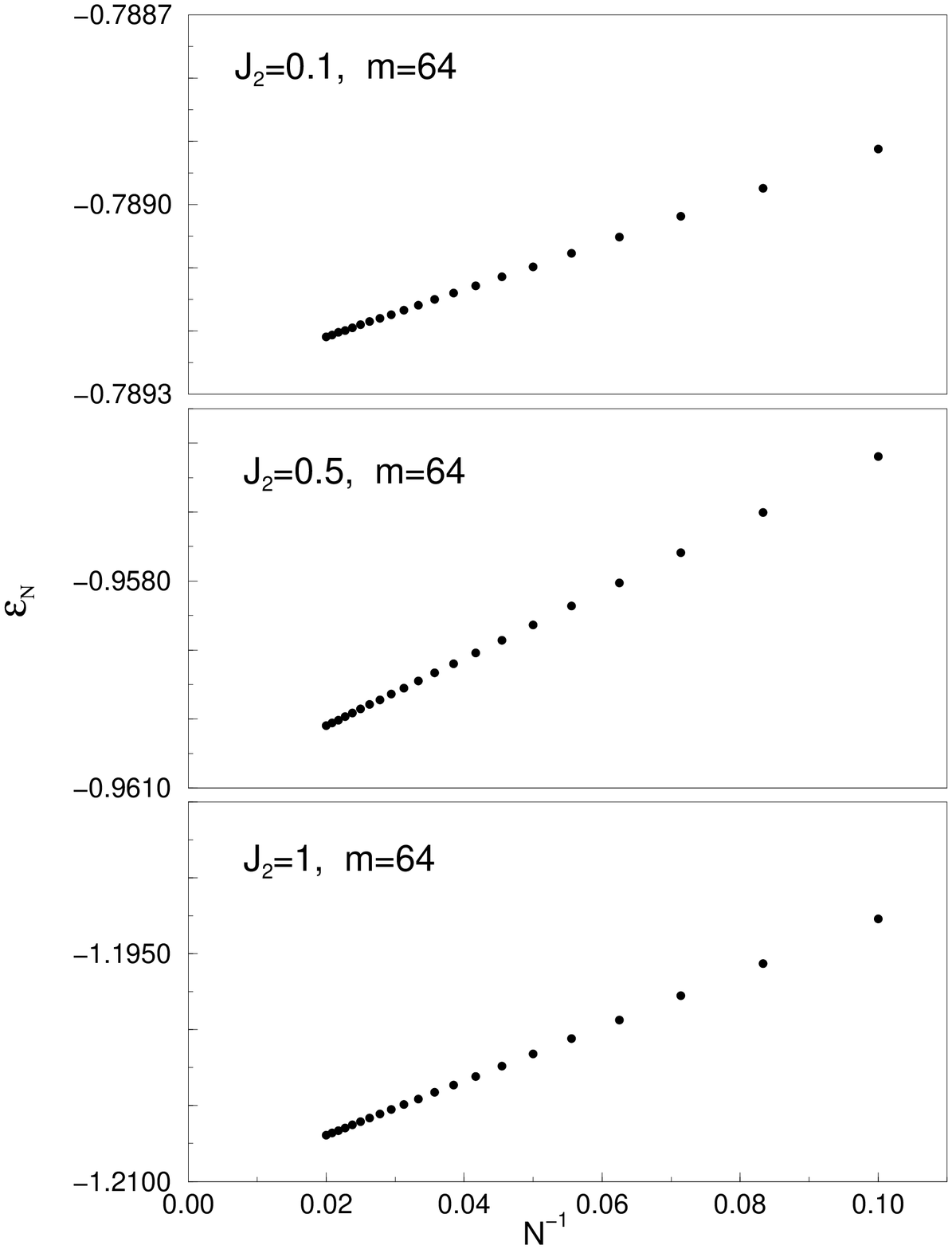,bbllx=50,bblly=0,bburx=500,bbury=800,height=20cm}
\end{center}
\vspace*{-1cm}
\centerline{Fig. 3}
\label{fig3}
\end{figure}

\begin{figure}[hp]
\begin{center}
\epsfig{figure=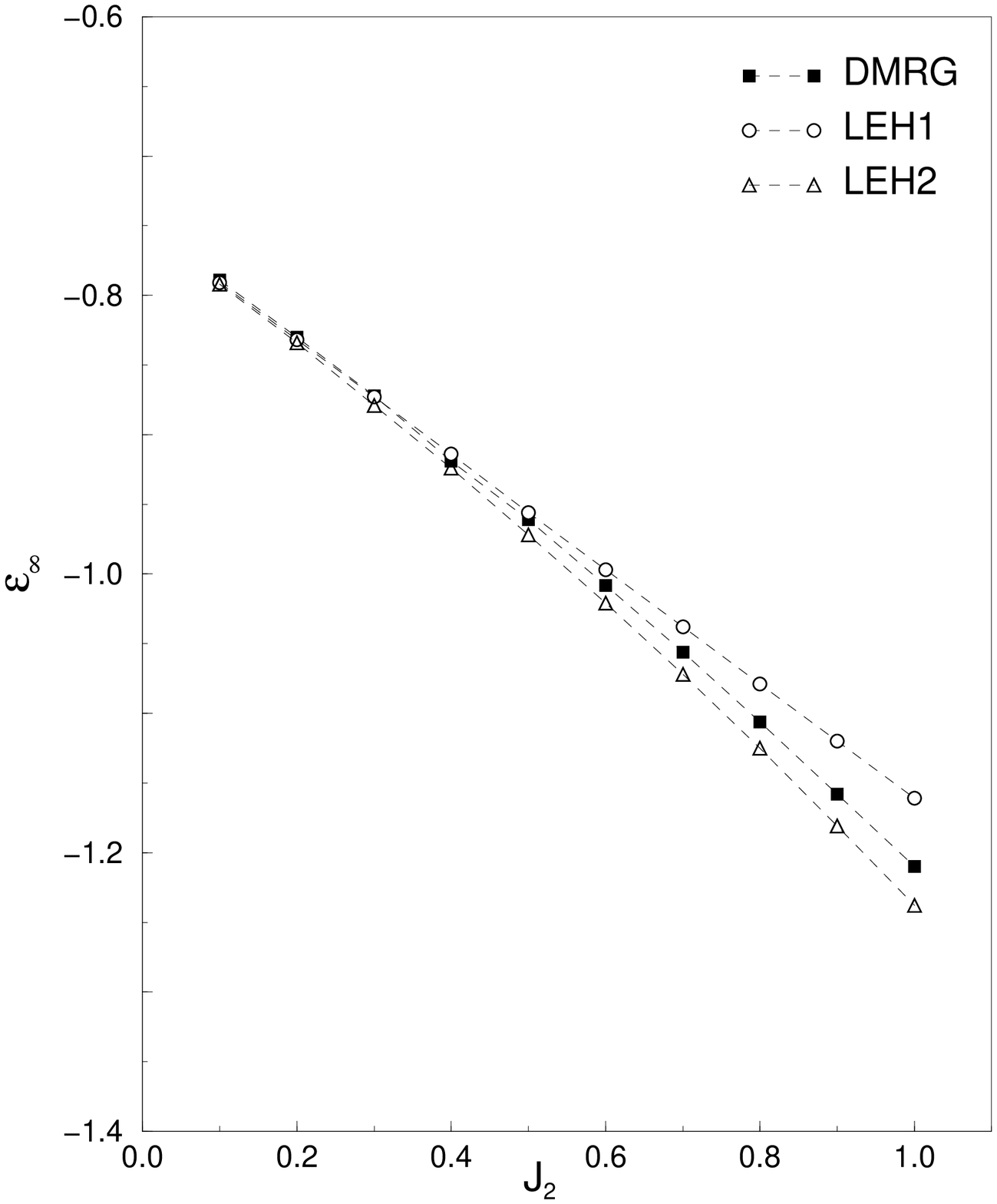,bbllx=50,bblly=0,bburx=500,bbury=800,height=20cm}
\end{center}
\vspace*{-1cm}
\centerline{Fig. 4}
\label{fig4}
\end{figure}

\begin{figure}[hp]
\begin{center}
\epsfig{figure=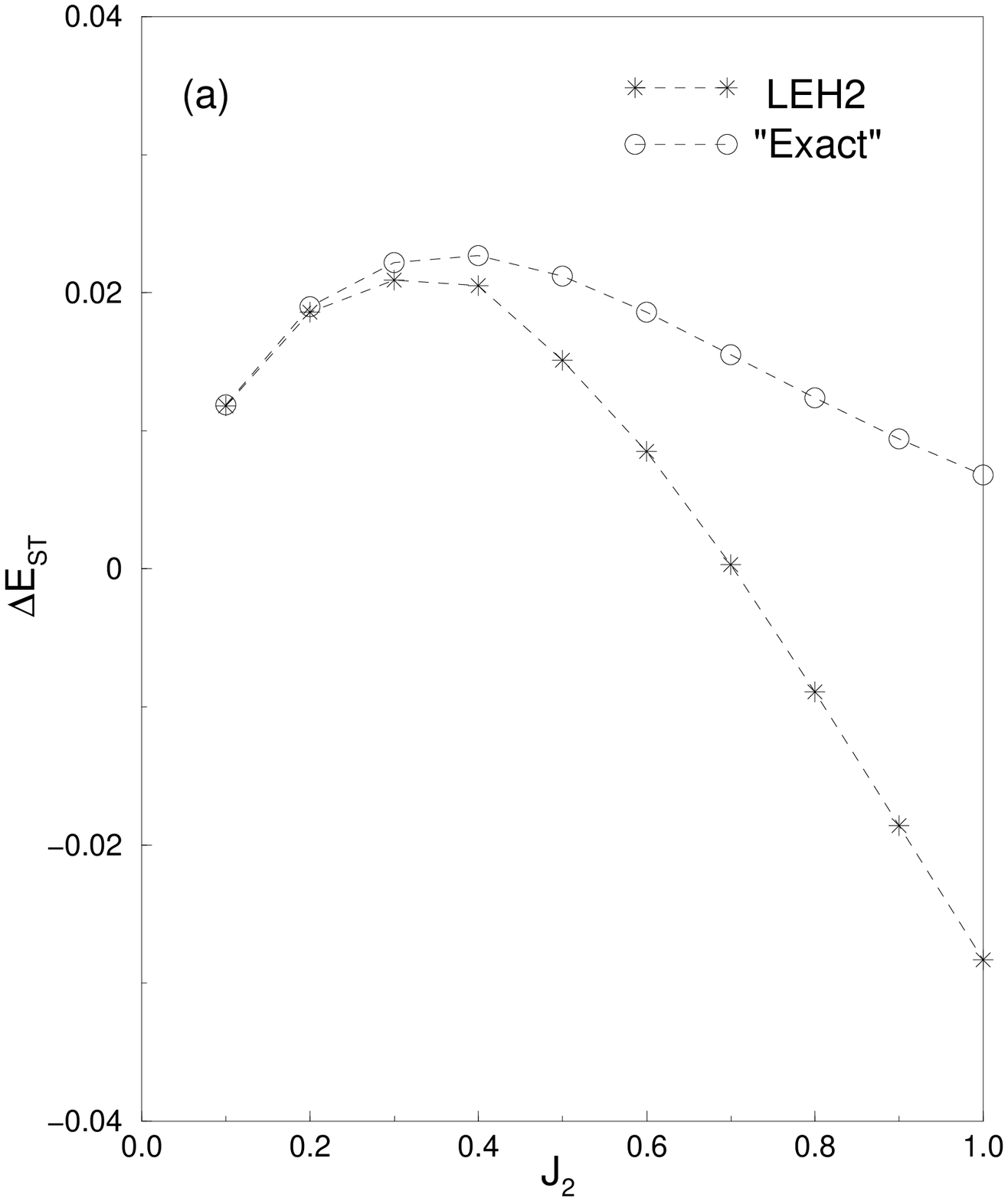,bbllx=50,bblly=0,bburx=500,bbury=800,height=20cm}
\end{center}
\vspace*{-1cm}
\centerline{Fig. 5 (a)}
\newpage
\begin{center}
\epsfig{figure=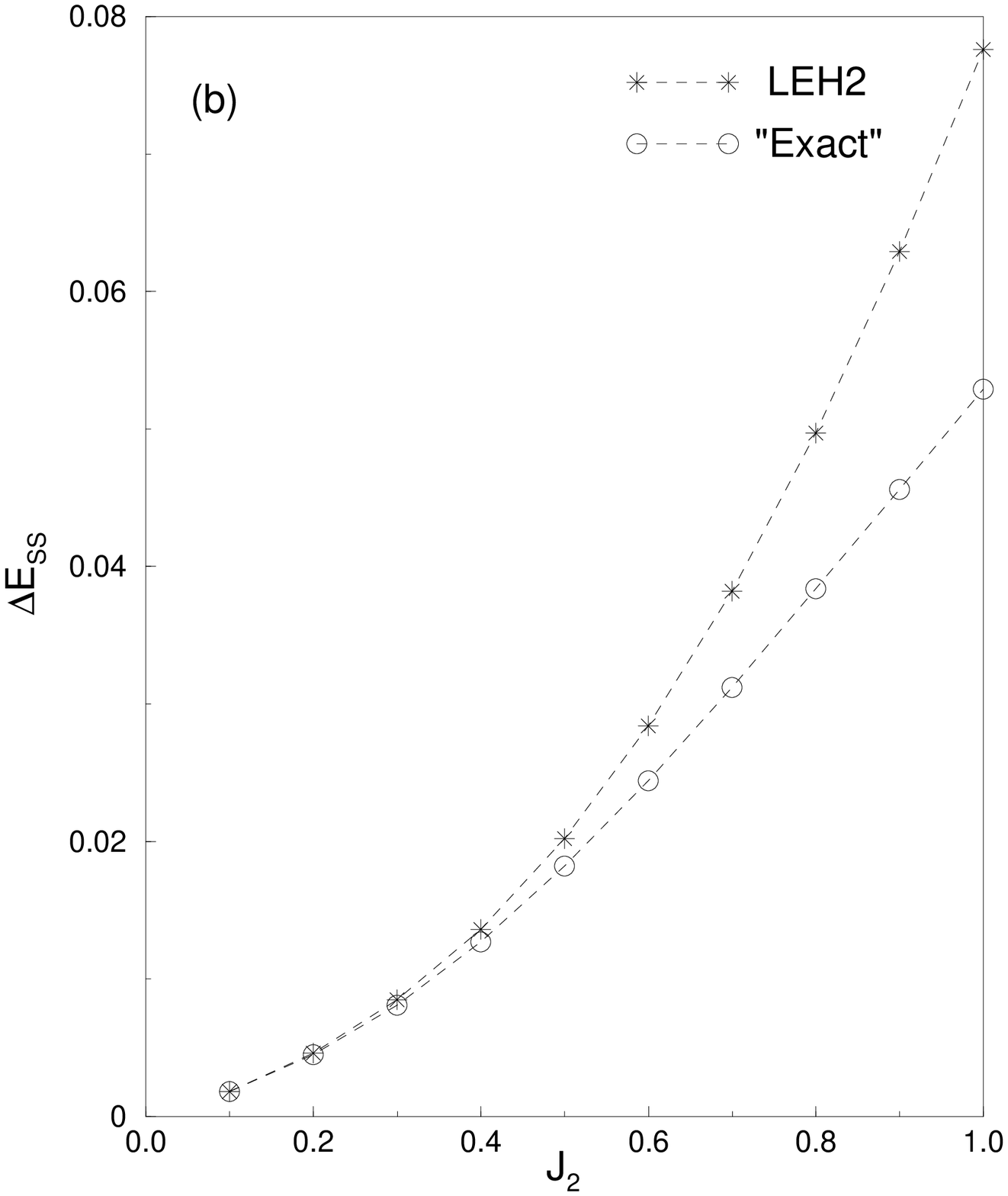,bbllx=50,bblly=0,bburx=500,bbury=800,height=20cm}
\end{center}
\vspace*{-1cm}
\centerline{Fig. 5 (b)}
\label{fig5}
\end{figure}

\begin{figure}[hp]
\begin{center}
\epsfig{figure=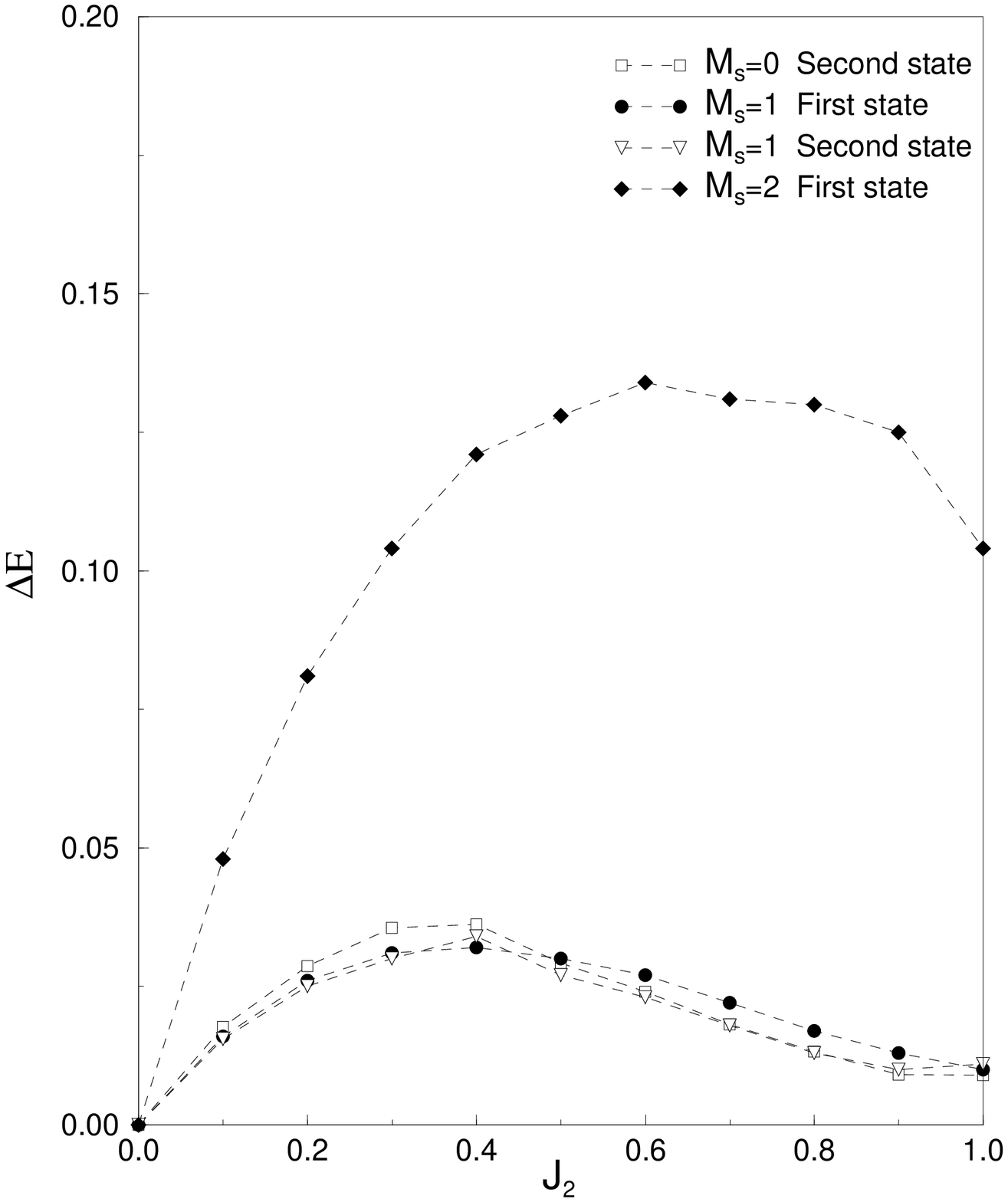,bbllx=50,bblly=0,bburx=500,bbury=800,height=20cm}
\end{center}
\vspace*{-1cm}
\centerline{Fig. 6}
\label{fig6}
\end{figure}

\begin{figure}[hp]
\begin{center}
\epsfig{figure=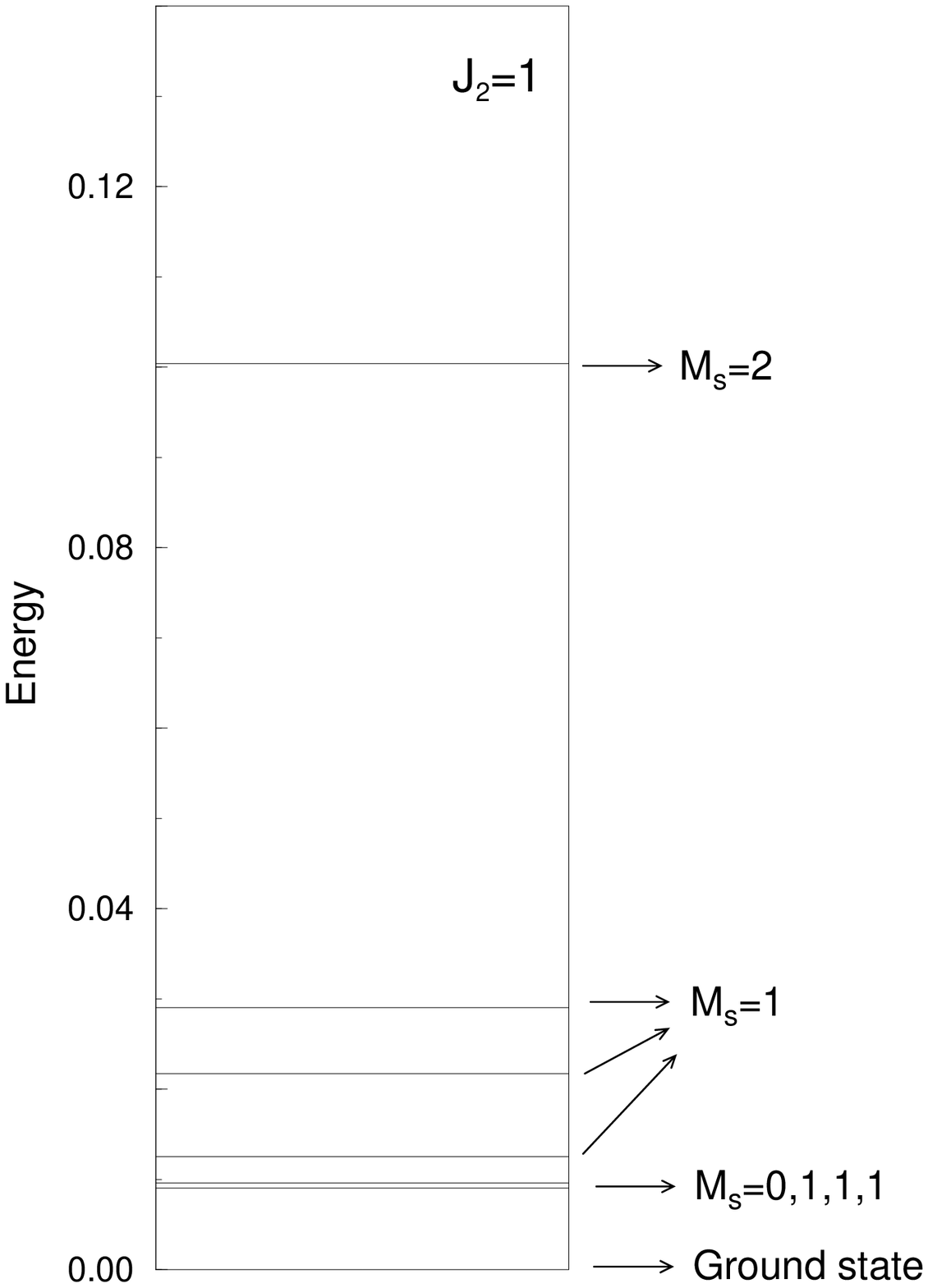,bbllx=50,bblly=0,bburx=500,bbury=800,height=20cm}
\end{center}
\vspace*{-1cm}
\centerline{Fig. 7}
\label{fig7}
\end{figure}

\begin{figure}[hp]
\begin{center}
\epsfig{figure=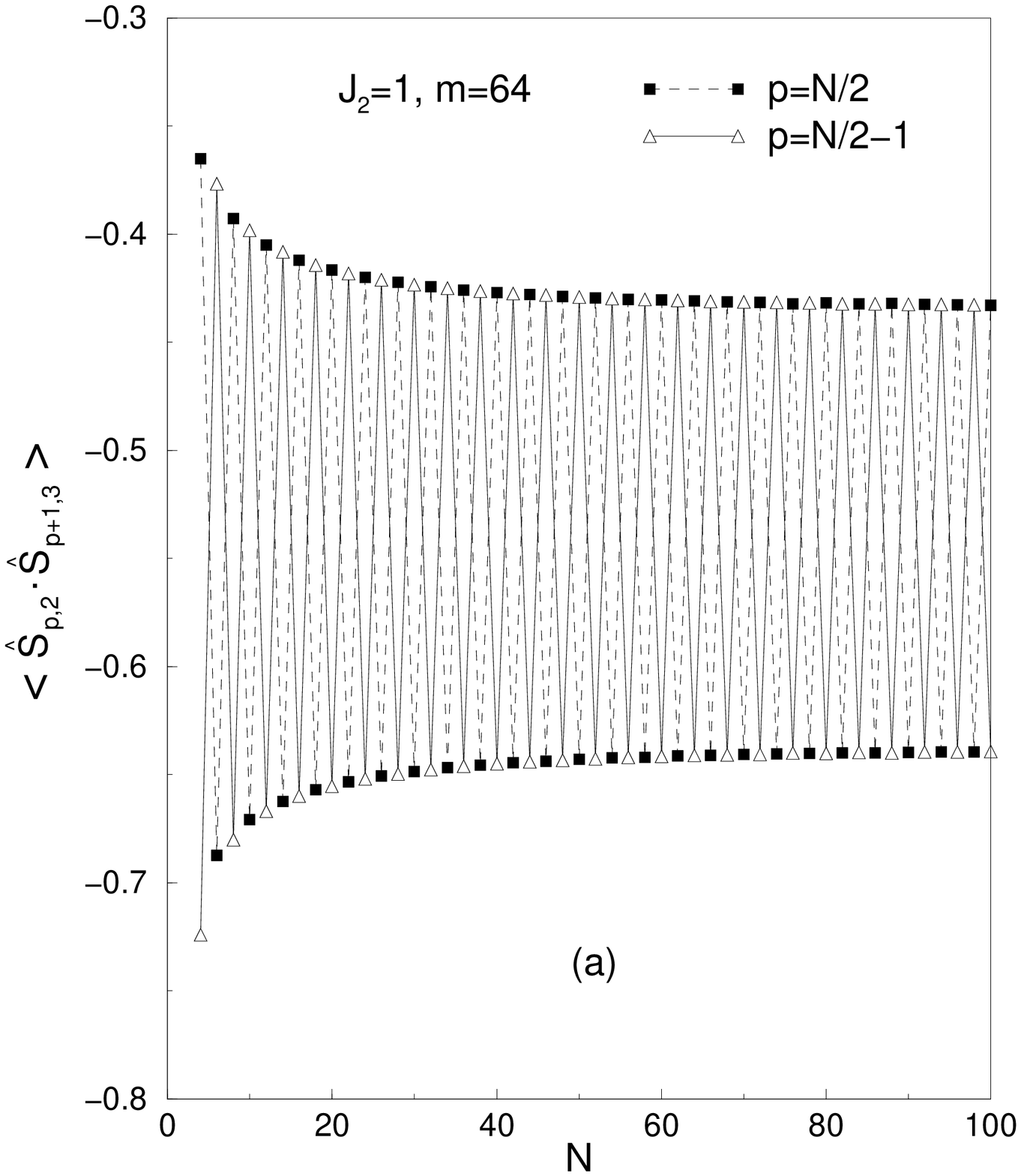,bbllx=50,bblly=0,bburx=500,bbury=800,height=20cm}
\end{center}
\vspace*{-1cm}
\centerline{Fig. 8 (a)}
\newpage
\begin{center}
\epsfig{figure=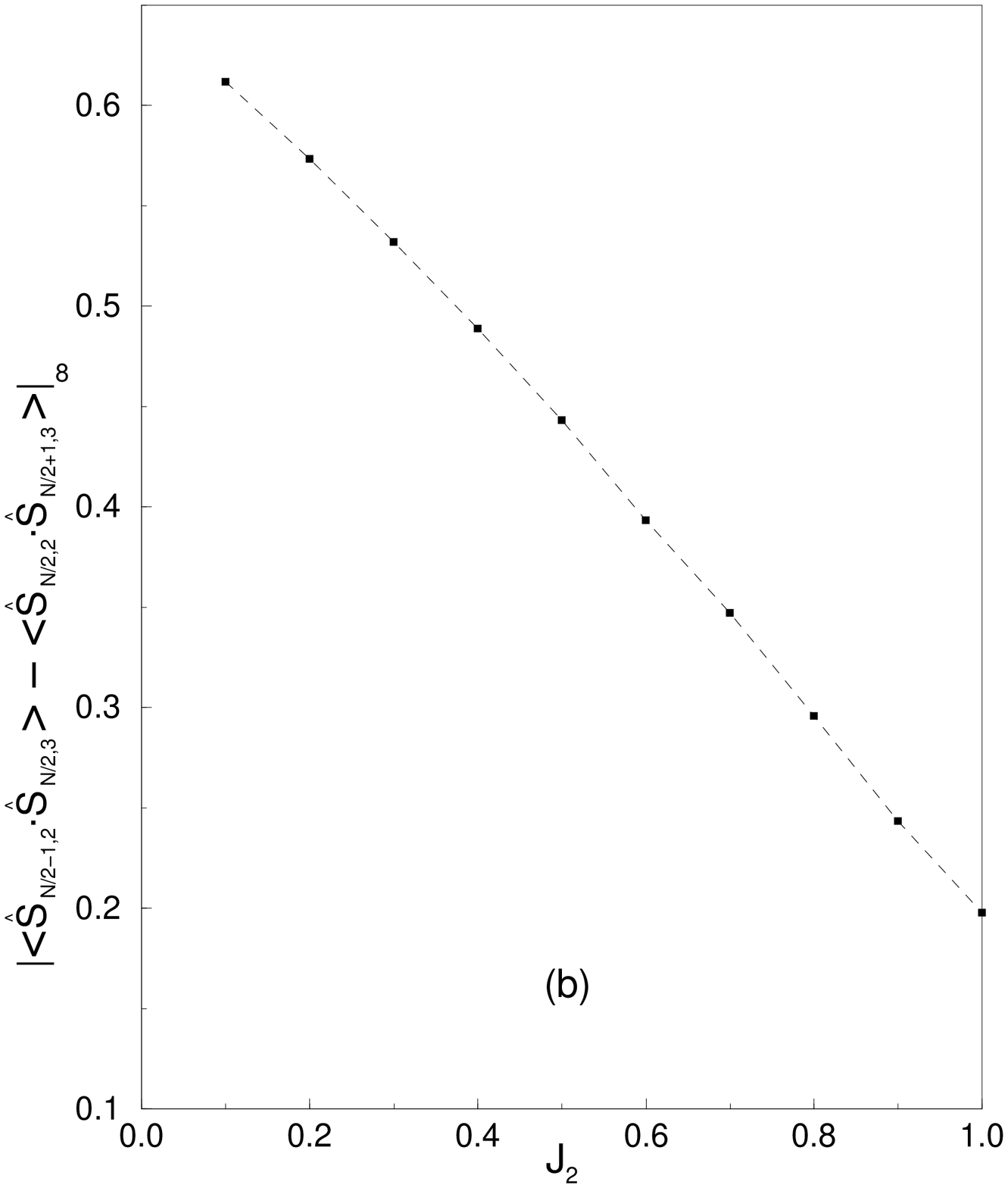,bbllx=50,bblly=0,bburx=500,bbury=800,height=20cm}
\end{center}
\vspace*{-1cm}
\centerline{Fig. 8 (b)}
\vspace*{1cm}
\label{fig8}
\end{figure}

\begin{figure}[hp]
\begin{center}
\epsfig{figure=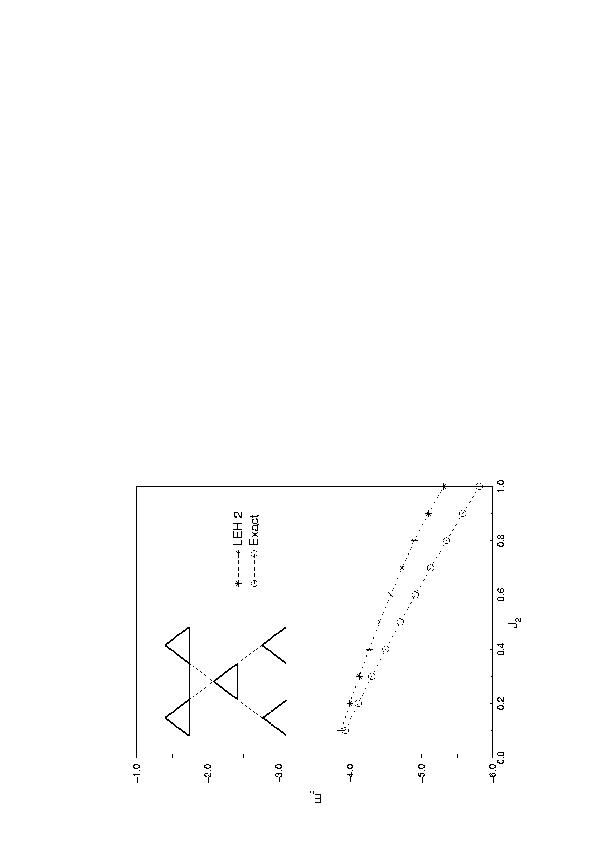,bbllx=50,bblly=-100,bburx=500,bbury=700,height=20cm,
angle=-90}
\end{center}
\vspace*{2cm}
\centerline{Fig. 9}
\label{fig9}
\end{figure}

\end{document}